# Multi-mode conversion via two-dimensional refractive-index perturbation on a silicon waveguide

Chunhui Yao, Zhen Wang, Hongwei Wang, Yu He, Yong Zhang & Yikai Su*

State Key Laboratory of Advanced Optical Communication Systems and Networks, Department of Electronic Engineering, Shanghai Jiao Tong University, Shanghai 200240, China

**Abstract**

Mode-division multiplexing offers a promising solution to increase the data capacity for optical communications. Waveguide mode conversion is essential for on-chip mode-division multiplexing. Previously reported mode converters have been limited to the conversion from a fundamental mode to one particular high-order mode. It is challenging to simultaneously satisfy the phase matching conditions during multiple mode conversion processes. Here, we propose a scalable design method that overcomes this limitation and realizes the simultaneous conversion of multiple modes via an all-dielectric two-dimensional metastructure on a silicon waveguide by shallow etching with hexagonal patterns. As an example, we experimentally demonstrate a multi-mode converter that simultaneously converts the $TE_i$ modes to the $TE_{i+3}$ ($i$ = 0, 1, 2) modes. The length of the multi-mode converter is 16.2 μm. The $TE_0$-$TE_3$, $TE_1$-$TE_4$, and $TE_2$-$TE_5$ mode conversions exhibit low insertion losses (0.4 to 1.0 dB) and reasonable crosstalk values (−14.1 to −16.5 dB) at 1538 nm.

___________________________________
E-mails: yikaisu@sjtu.edu.cn

**Introduction**

Generation and manipulation of optical modes are of broad interests to the photonics community. In particular, space-division-multiplexing (SDM) or mode-division multiplexing (MDM) shows great potential to increase the data capacity in optical communication systems[1–6]. On-chip silicon MDM offers compact footprint and compatibility with complementary metal-oxide-semiconductor (CMOS) technologies[7–9]. In the past few years, various on-chip mode converters and multiplexers have been proposed based on phase matching[10–13], beam shaping[14–16] and coherent scattering[17–19]. A promising technique is to design mode converters using metastructures to control the effective refractive-index distribution, thereby manipulating the propagation and conversion of the waveguide modes[20–27]. Previously proposed metastructures on waveguides, such as graded index co-directional grating with periodic index variation along the propagation direction[20], phase-gradient metasurface structures consisting of nanoantenna arrays[21] and all-dielectric metasurface structure with tilted subwavelength periodic perturbations[22], were capable of converting one fundamental mode to a selective higher order mode.

A common issue of the reported mode converters is that the mode conversion is limited to one pair of modes. It would be much more attractive to realize simultaneous conversion processes from multiple low-order modes to high-order modes in a multimode waveguide, in analogy to the multi-wavelength conversion in the wavelength domain[28-30]. Such a multi-mode converter should feature compact footprint, wide operation bandwidth, and low insertion loss. The key challenge of realizing multi-mode conversion is to achieve equally high conversion efficiencies for multiple mode pairs. Due to the significant discrepancy of the phase matching conditions between different pairs of modes, multi-mode conversion is difficult to realize using conventional phase matching techniques based on periodic perturbations[20,22].

In this paper, we propose and demonstrate an efficient low-loss multi-mode converter using a quasi two-dimensional metastructure on a silicon waveguide. Shallow hexagonal trenches etched on the silicon waveguide provide sufficient refractive-index perturbations and introduce low excess losses. A segmented index profile in the transverse direction is designed based on coupled mode theory to maximize target mode coupling coefficients, while a quasi-periodic refractive-index variation along the propagation direction is optimized by particle swarm optimization (PSO) algorithm to achieve approximately equal conversion efficiencies for multiple pairs of modes. As a proof-of-concept experiment, we demonstrate a multi-mode converter that can realize the simultaneous conversions from $TE_i$ modes to $TE_{i+3}$ ($i$ = 0, 1, 2) modes. The length of the multi-mode converter is 16.2 μm. For the $TE_0$-to-$TE_3$, $TE_1$-to-$TE_4$ and $TE_2$-to-$TE_5$ mode conversions, the measured insertion losses are 0.4 dB, 1.0 dB and 0.5 dB, respectively, and the corresponding crosstalk values are below −15.5 dB, −16.5 dB and −14.1 dB, respectively, at 1538 nm. The proposed multi-mode converter can be scaled to realize simultaneous mode conversions from multiple low-order modes to arbitrary high-order modes.

**Results**

**Theoretical principle for multi-mode conversion.** The propagation of optical field in a perturbed dielectric structure can be approximately described by the coupled mode theory. Attributed to the index perturbations on a silicon multimode waveguide, energy can be coupled from one waveguide mode to other modes, and the amplitude of each mode along the propagation direction $z$ can be determined by a set of differential equations[20,22,31]

$$-\frac{\partial A_p}{\partial z} = \sum_{q=0(q \neq p)}^{h} j\kappa_{pq} A_q e^{j(\beta_p - \beta_q)z}, \quad \left(p = 0, 1, ..., h\right), \tag{1}$$

where $A_p$ and $A_q$ are the amplitudes of the waveguide modes $p$ and $q$, and $\beta_p$ and $\beta_q$ are their propagation constants, respectively. $h$ denotes the order of the highest order mode in the multimode waveguide. $\kappa_{pq}$ represents the mode coupling coefficient between the waveguide modes $p$ and $q$, which can be defined as[20,22,31]

$$\kappa_{pq}(z) = \frac{\omega}{4} \iint_S \Delta\varepsilon(x, y, z) \cdot E_p^*(x, y) \cdot E_q(x, y) dxdy, \tag{2}$$

where the integration region $S$ is the cross section of the silicon waveguide, $E_p(x, y)$ and $E_q(x, y)$ are the electric field profiles of the waveguide modes $p$ and $q$ in the transverse cross section, respectively, and $\Delta\varepsilon(x, y, z)$ represents the refractive-index perturbation on the waveguide. According to Eq. 2, $\kappa_{pq}$ is proportional to the spatial integral of the perturbation $\Delta\varepsilon(x, y, z)$ with $E_p(x, y)$ and $E_q(x, y)$, which we refer to as the electric field overlap.

There are two requirements needed for the multi-mode conversion based on Eq. 1. The first one is to achieve large values of the mode coupling coefficients between the incident modes and the target modes. The second one is to satisfy the phase matching condition along the propagation direction, which compensates the propagation constant mismatch $\Delta\beta$ (i.e. $\beta_p - \beta_q$) of the oscillating exponential term in Eq. 1. The phase matching condition for the mode coupling between modes $p$ and $q$ can be inferred as[20,22,31]

$$\delta_{pq} = \frac{2\pi}{\beta_p - \beta_q}, \tag{3}$$

where $\delta_{pq}$ represents the period of the index perturbation. The discrepancy of $\Delta\beta$ between different pairs of modes leads to various perturbation periods and coupling lengths. For instance, the calculated perturbation periods $\delta_{i(i+3)}$ ($i$ = 0, 1, 2) for the TE$_i$-to-TE$_{i+3}$ mode couplings are about 5.8 μm, 4.7 μm and 3.8 μm, respectively. Therefore, it is difficult to simultaneously satisfy the phase matching conditions for multiple pairs of modes using the periodic index perturbations.

To meet these two requirements, our solution is to employ two-dimensional refractive-index perturbations composed of shallow-etched trenches on a multimode waveguide. Along the transverse direction, following Eq. 2, the perturbations are introduced in the regions where the electric field profiles of the mode pairs of interest overlap constructively to obtain large values of the mode coupling coefficients. Along the propagation direction, shallow-etched trenches with different lengths form quasi-periodic perturbations. By properly designing the length of each trench, the phase matching between multiple pairs of modes can be achieved simultaneously.

**Device design and numerical simulation.** We first use simple rectangular shallow-etched trenches on a silicon multimode waveguide to form the two-dimensional refractive-index perturbations and realize the multi-mode conversions from TE$_i$ modes to TE$_{i+3}$ ($i$ = 0, 1, 2) modes, as shown in Fig. 1a. Along the transverse direction, the widths of the trenches ($W_1$ to $W_4$) are properly designed to achieve large values of the mode coupling coefficients $\kappa_{03}$, $\kappa_{14}$ and $\kappa_{25}$. Figure 1b shows the simulated electric field profiles $E$ ($x$, $y$) of multiple mode pairs (TE$_0$ / TE$_3$, TE$_1$ / TE$_4$ and TE$_2$ / TE$_5$) in the transverse section, and the black-and-white-hatched regions highlight the field overlapping regions between the TE$_0$ and TE$_3$ modes, the TE$_1$ and TE$_4$ modes, the TE$_2$ and TE$_5$ modes, respectively. Based on Eq. 2, by introducing the trenches to these overlapping regions, the maximum positive values of $\kappa_{03}$, $\kappa_{14}$ and $\kappa_{25}$ can be achieved, respectively. It can be noted that the field overlap regions between multiple mode pairs are slightly different, thus the trenches are introduced with trade-off widths that can cover the vast majority of them, as shown by the red-dashed areas in Fig. 1b. Benefitting from the suitable trench placement, sufficiently large values of $\kappa_{03}$, $\kappa_{14}$ and $\kappa_{25}$ can be simultaneously achieved, which lead to high mode conversion efficiencies and short coupling lengths. In this design, the width of the silicon waveguide is 3.0 μm, and the calculated widths of the trenches are: $W_1 = W_4 = 0.46$ μm and $W_2 = W_3 = 1.04$ μm. After propagation over $\delta_{pq}/2$, the incident mode and the target mode are out of phase; thus changing the value of $\kappa_{pq}$ from positive to negative is required to ensure that the input mode can always constructively contribute to the conversion to the target mode[20,22,31]. Therefore, the trenches should be arranged in a checkered pattern alternating along the propagation direction for the reversals of the signs of the mode coupling coefficients, as shown in Fig. 1a. Note that larger mode coupling coefficients can be achieved with deeper etching of the shallow trenches, at the cost of higher excess losses. In our case, the etching depth is chosen as 50 nm to introduce sufficient index perturbation and negligible losses.

Along the propagation direction, we use the PSO algorithm[32] to optimize the lengths of the rectangular trenches ($L_1$ to $L_6$) in order to achieve high conversion efficiencies between the $TE_i$ and $TE_{i+3}$ ($i = 0, 1, 2$) modes. Regarding the transmissions of the target modes ($TE_3$ / $TE_4$ / $TE_5$) as the target of optimization, the PSO algorithm is expected to find the optimal set of the lengths via adequate iterative calculations. Detailed parameters and simulated performance are provided in Supplementary Table 1. The simulated insertion losses are ~ 1 dB for the $TE_i$-to-$TE_{i+3}$ ($i = 0, 1, 2$) multi-mode conversion. The considerable insertion losses can be mainly attributed to the Fresnel reflection at the boundaries of the rectangular trenches caused by the high refractive-index contrast between the silicon core and the cladding.

A viable way to reduce the losses caused by the reflection is to introduce taper structures at the interface of the etched region and the silicon waveguide. We add two symmetrical tapers with the same length $T$ at the two sides of the rectangular trench to form a hexagon structure, as shown in Fig. 1c. Figure 1d,e depicts the top view and three-dimensional (3D) view of the proposed $TE_i$-to-$TE_{i+3}$ ($i = 0, 1, 2$) multi-mode converter with the two-dimensional metastructure composed of the hexagonal trenches, respectively. Along the transverse direction, the widths of the hexagonal trenches are of the identical magnitude with those of the rectangular ones. Since the introducing of the taper structure, the optimal lengths for the hexagonal trenches are changed compared to the rectangular trenches. So, we apply PSO algorithm again to search the optimal length parameters ($T_1$ to $T_6$ and $L_1$ to $L_6$). To simplify the calculation process, we assume the ratio $R = T_n / L_n$ ($n = 1, 2, …, 6$) to be a constant, thus limiting the parameter set ($T_1$ to $T_6$ and $R$) in the optimization. Table 1 provides the optimized structure parameters. By employing the hexagonal trenches, the simulated insertion losses of the $TE_i$-to-$TE_{i+3}$ ($i = 0, 1, 2$) multi-mode conversion are optimized to be lower than 0.5 dB.

3D finite-difference time-domain (FDTD) method is used for optical simulations. Figure 2a-c shows the calculated mode coupling coefficients and mode evolution processes for multiple mode pairs and the simulated optical field distributions in the multi-mode converter. When the $TE_0$ / $TE_1$ / $TE_2$ mode light is launched into the input port, the light is coupled and converted to the $TE_3$ / $TE_4$ / $TE_5$ mode, respectively. Attributed to index perturbation along the propagation direction, the mode coupling coefficient $\kappa_{i(i+3)}$ behave like a clipped sinusoidal function, which can help the mode conversion process. The phase matching between multiple pairs of mode are eventually achieved through the two-dimensional perturbation structure on the waveguide, thus the input modes are gradually converted to the target modes and reach similarly high mode purities of ~ 94%, where the mode purity is defined as the ratio of the transmitted target mode power to the total transmitted power.

By varying the etching depths and simultaneously changing the widths and the lengths of the hexagon trenches, we investigate the fabrication tolerance of our proposed multi-mode converter by simulations. As shown in Fig. 3, when the etching depth varies from 45 nm to 55 nm, or the width and length deviation of each hexagon varies from −150 nm to +150 nm, the insertion losses and the crosstalk values of the multi-mode converter still remain < 1 dB and < −10 dB, respectively. The simulated results indicate that the proposed devices feature high tolerance to structural deformation. The tip of the taper structure may be difficult to fabricate due to the small feature size, so we use a trapezoid taper with a tip size of 50 nm to replace the triangle taper to mimic the shape deformation. The simulated results show negligible degradation of the device performance, which indicates that the proposed devices are tolerant to the deformation of the tip.

To verify the scalability of our proposed design method, we also design a $TE_i$-to-$TE_{i+2}$ ($i$ = 0, 1) multi-mode converter and a $TE_i$-to-$TE_{i+4}$ ($i$ = 0, 1, 2) multi-mode converter employing the shallow-etched hexagonal trenches. The simulated electric field ($E_y$) distributions are shown in Supplementary Fig. 1 and Fig. 2, respectively. For the $TE_i$-to-$TE_{i+2}$ ($i$ = 0, 1) multi-mode conversion, the simulated insertion losses are 0.48 dB and 0.44 dB, and the crosstalk values are below −15.2 dB and −12.8 dB, respectively. For the $TE_i$-to-$TE_{i+4}$ ($i$ = 0, 1, 2) multi-mode conversion, the simulated insertion losses are 0.54 dB, 0.71 dB and 0.67 dB, and the crosstalk values are below −14.4 dB, −16.0 dB and −12.4 dB, respectively. Detailed device parameters and simulation results are provided in the Supplementary Table 2 and Table 3, respectively.

**Optical performance of the fabricated devices.** Figure 4a shows an optical microscope photo of a fabricated $TE_i$-to-$TE_{i+3}$ ($i$ = 0, 1, 2) multi-mode converter. Grating couplers were used to vertically couple the light into/out of the chip, with a coupling loss of ~ 7.9 dB/port. A 3-channel multiplexer ($TE_0$ ~ $TE_2$, with input ports $I_0$ ~ $I_2$) and a 6-channel de-multiplexer ($TE_0$ ~ $TE_5$, with output ports $O_0$~$O_5$) based on asymmetrical directional couplers (DCs)[10] are cascaded before and after the fabricated multi-mode converter to characterize the performance. The injected $TE_0$ modes from the input port $I_i$ ($i$ = 0, 1, 2) are multiplexed to the $TE_i$ ($i$ = 0, 1, 2) mode in the bus waveguide by the 3-channel multiplexer and then launched into the multi-mode converter. After mode conversions, the 6-channel de-multiplexer recovers all the output mode signals to the $TE_0$ modes for the measurements. Detailed parameters for the mode (de)multiplexer are provided in Supplementary Table 4. The measured insertion losses of the fabricated $TE_1$~$TE_5$ modes (de)multiplexers are 5.6, 6.9, 5.0, 5.2 and 5.4 dB at $\lambda$ = 1538 nm, respectively. Figure 4b,c shows the scanning electron microscope (SEM) images of a fabricated $TE_i$-to-$TE_{i+3}$ ($i$ = 0, 1, 2) multi-mode converter.

Figure 5a-c presents the measured transmission responses of all the output ports ($O_0$ ~ $O_5$) when the light is launched into the input ports ($I_0$ ~ $I_2$), respectively. The transmission spectra were normalized to identical grating couplers and the mode (de)multiplexers fabricated on the same wafer. For the $TE_0$-to-$TE_3$, $TE_1$-to-$TE_4$ and $TE_2$-to-$TE_5$ mode conversions, the measured insertion losses are 0.4 dB, 1.0 dB and 0.5 dB, and the crosstalk values are −15.5 dB, −16.5 dB and −14.1 dB at 1538 nm, respectively. In the wavelength range of 1525 nm to 1550 nm, the overall insertion losses are lower than 3.5 dB, and the crosstalk values are below −11.5 dB. Simulated transmission spectra are also presented for comparison, as shown by the dashed curves in Fig. 5a-c. The fluctuation of the measured transmission spectra and the narrow operation bandwidth of the multi-mode converter may be attributed to the poor (de)multiplexing performance of the conventional DC-based

(de)multiplexers, which is sensitive to the fabrication errors for high-order mode coupling. To characterize the performance of the multi-mode converter more accurately, a possible solution is to replace the conventional DCs with subwavelength-grating-based DCs[6].

**Discussion**

In conclusion, we have proposed and experimentally demonstrated a scalable design method to achieve the simultaneous conversion of an arbitrary group of modes using a two-dimensional metastructure on a silicon waveguide. Shallow-etched hexagonal trenches are used to control the spatial distribution of the effective refractive-index. A segmented refractive-index profile along the transverse direction and quasi-periodic index perturbations along the propagation direction are applied to form two-dimensional index perturbations and achieve equally high conversion efficiencies for multiple pairs of modes. As an example, an efficient low-loss multi-mode converter is demonstrated to achieve the simultaneous mode conversions from the $TE_i$ mode to the $TE_{i+3}$ mode ($i$ = 0, 1, 2). The length of the multi-mode converter is 16.2 μm. The measured insertion losses of the $TE_0$-to-$TE_3$, $TE_1$-to-$TE_4$ and $TE_2$-to-$TE_5$ mode conversions are 0.4 dB, 1.0 dB and 0.5 dB, and the crosstalk values are −15.5 dB, −16.5 dB and −14.1 dB at 1538 nm, respectively.

Enabled by the two-dimensional quasi-periodic index perturbations, the proposed design method overcomes the limitation of single pair mode conversion and provides a general approach to simultaneously manipulating multiple on-chip waveguide modes. The demonstrated multi-mode conversion device may provide great flexibility in the design and implementation of integrated MDM communication systems and advanced optical information processing systems.

## Methods

**Device fabrication and characterization.** The proposed $TE_i$-to-$TE_{i+3}$ ($i$ = 0, 1, 2) multi-mode converters were fabricated on a silicon-on-insulator (SOI) wafer (220 nm thick silicon on 3000 nm thick silica). Grating couplers, silicon waveguides, and shallow-etched hexagonal trenches were patterned and etched by E-beam lithography (Vistec EBPG 5200$^+$) and inductively coupled plasma etching (SPTS DRIE-I), respectively. The devices were characterized using a tunable continuous wave laser (Keysight 81960A) and an optical power meter (Keysight N7744A).

asymmetric graded index photonic structures. *Opt. Lett.* **38**, 220–222 (2013).

17. Lejman, M. *et al*. Ultrafast acousto-optic mode conversion in optically birefringent ferroelectrics. *Nat. Commun*. **7**, 10 (2016).

18. Liu, V., Miller, D. A. & Fan, S. Ultra-compact photonic crystal waveguide spatial mode converter and its connection to the optical diode effect. *Opt. Express* **20**, 28388–28397 (2012).

19. Frandsen, L. H. Topology optimized mode conversion in a photonic crystal waveguide fabricated in silicon-on-insulator material. *Opt. Express* **22**, 8525–8532 (2014).

20. Ohana, D., Desiatov, B., Mazurski, N. & Levy, U. Dielectric metasurface as a platform for spatial mode conversion in nanoscale waveguides. *Nano Lett*. **16**, 7956–7961 (2016).

21. Li, Z. *et al*. Controlling propagation and coupling of waveguide modes using phase-gradient metasurfaces. *Nat. Nanotechnol*. **12**, 675–683 (2017).

22. Wang, H. Compact silicon waveguide mode converter employing dielectric metasurface structure. *Adv. Opt. Mater.* **7**, 1801191–1801191 (2019).

23. Marcuse, D. Mode conversion caused by surface imperfections of a dielectric slab waveguide. *Bell Syst. Tech. J.* **48**, 3187–3215 (1969).

24. Lin, D., Fan, P., Hasman, E. & Brongersma, M. L. Dielectric gradient metasurface optical elements. *Science* **345**, 298–302 (2014).

25. Tong, X. C. *Functional metamaterials and metadevices* (Springer, 2018).

**Acknowledgements**

The authors would like to thank the Center for Advanced Electronic Materials and Devices (AEMD) of Shanghai Jiao Tong University for the support in device fabrications.

**Author contributions**

C.Y., Z.W. and H.W. proposed the device schematic. C.Y. performed numerical simulations. Y.Z. fabricated the devices. C.Y. conceived and carried out the experiments with Z.W.'s assistance. C.Y. wrote the manuscript with Y.Z., Y.H. and H.W.'s contribution. Y.S. supervised the project, provided valuable support and feedback, and revised the manuscript.

**Figures**

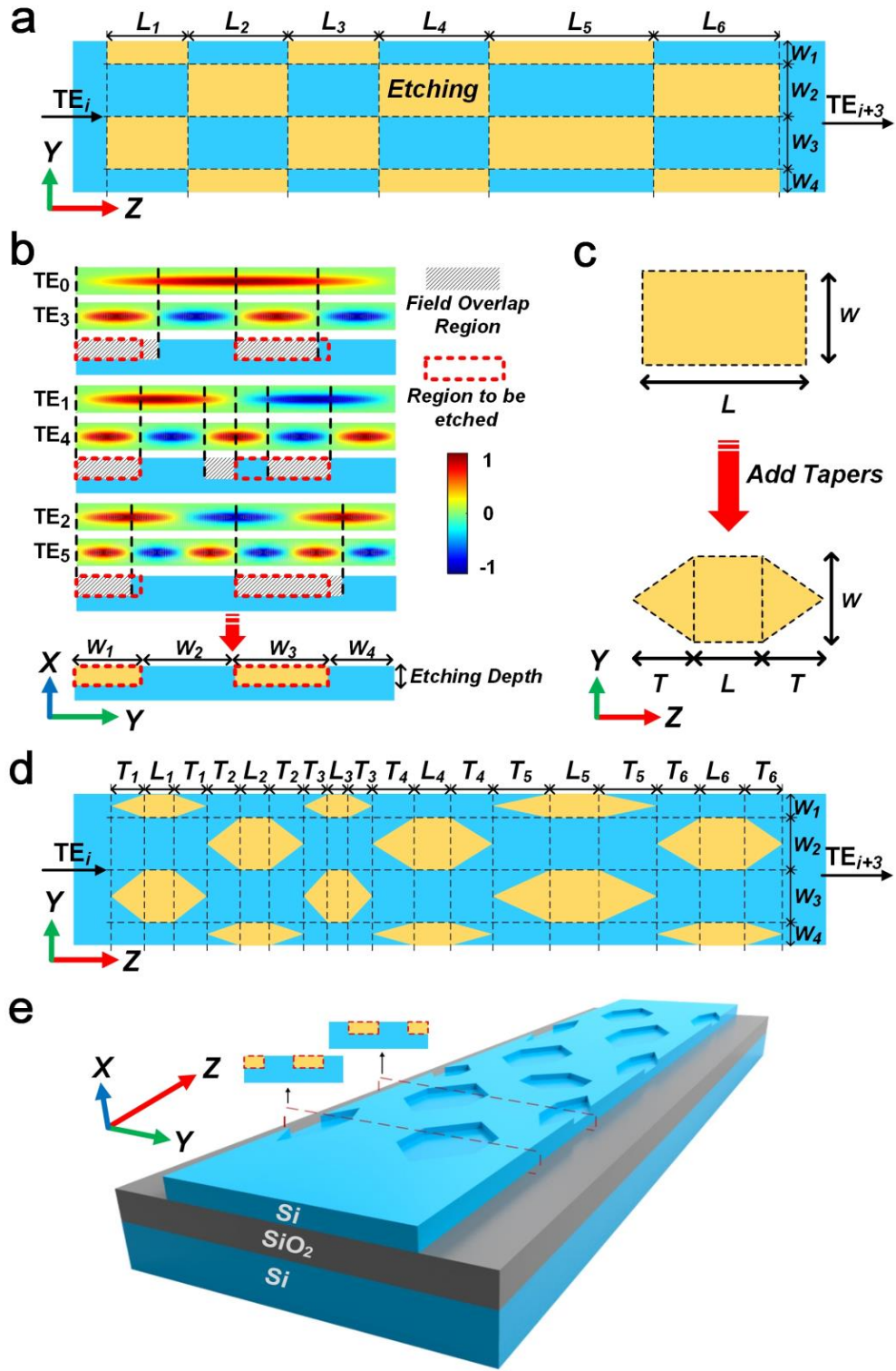

**Figure 1 | Multi-mode converters based on two-dimensional metastructure composed of rectangular or hexagonal trenches**. (**a**) Top view of the proposed TE$_i$-to-TE$_{i+3}$ ($i$ = 0, 1, 2) multi-mode converter with rectangular trenches. (**b**) Electric field profiles of the TE$_0$ / TE$_3$, TE$_1$ / TE$_4$ and TE$_2$ / TE$_5$ mode pairs in the transverse section. The black-and-white-hatched regions represent the field overlap regions between different pairs of modes in the transverse section, respectively. The trenches are introduced with trade-off widths to cover the vast majority of the overlap regions, as shown by the red-dashed regions. (**c**) Schematic of the rectangular trench and the hexagonal trench. (**d**) Top view of the proposed TE$_i$-to-TE$_{i+3}$ ($i$ = 0, 1, 2) multi-mode converter with hexagonal trenches. (**e**) 3D view of the proposed TE$_i$-to-TE$_{i+3}$ ($i$ = 0, 1, 2) multi-mode converter with hexagonal trenches. Insets show the cross sections of the waveguide.

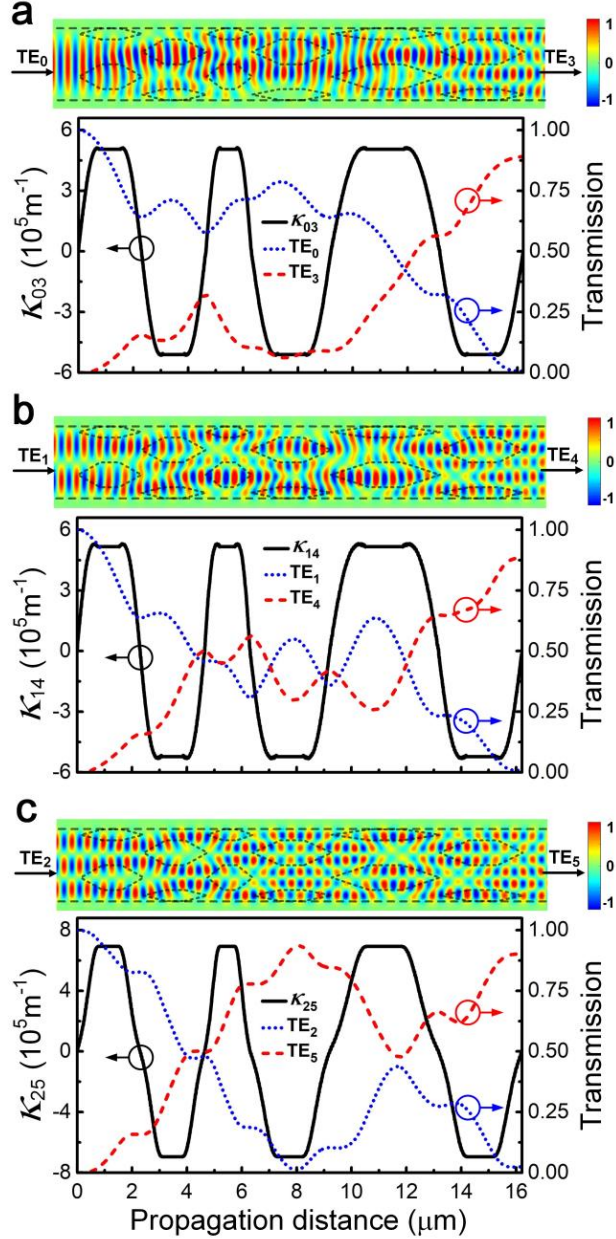

**Figure 2 | Simulation results of the multi-mode converter.** Simulated electric field ($E_y$) distribution, calculated mode coupling coefficients (black thick curve) and the transmissions of incident modes (blue dotted curve) and target mode (red dashed curve) along the propagation direction for the proposed $TE_i$-to-$TE_{i+3}$ ($i = 0, 1, 2$) multi-mode converter when the input light is (**a**) $TE_0$, (**b**) $TE_1$ and (**c**) $TE_2$ mode, respectively.

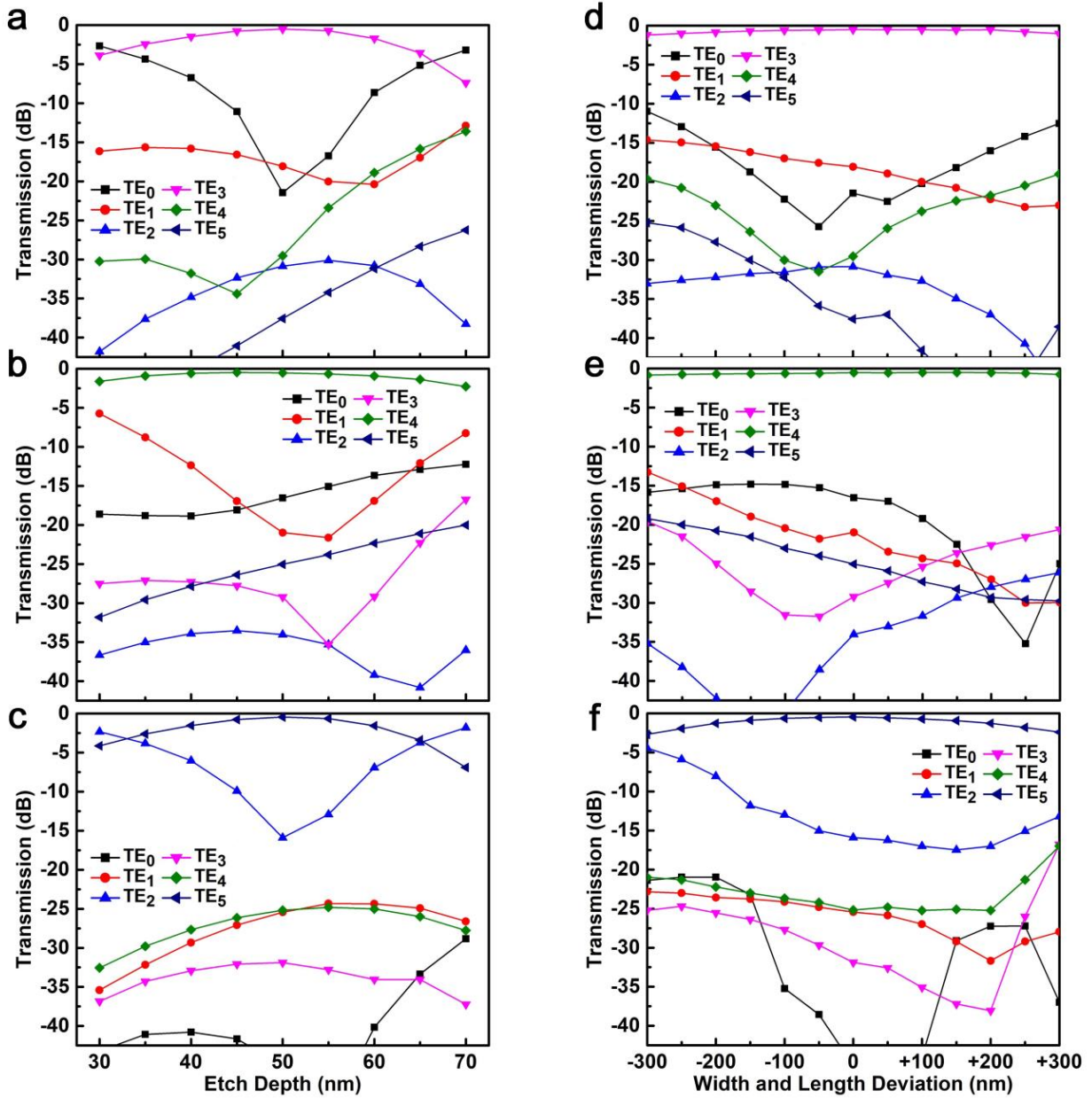

**Figure 3 | Tolerance study of the $TE_i$-to-$TE_{i+3}$ ($i$ = 0, 1, 2) multi-mode converter.** Simulated transmission responses by changing the etching depth for the (**a**) $TE_0$, (**b**) $TE_1$, and (**c**) $TE_2$ mode input, respectively. Simulated transmission responses by simultaneously changing the width and the length of each hexagon for the (**d**) $TE_0$, (**e**) $TE_1$, and (**f**) $TE_2$ mode input, respectively.

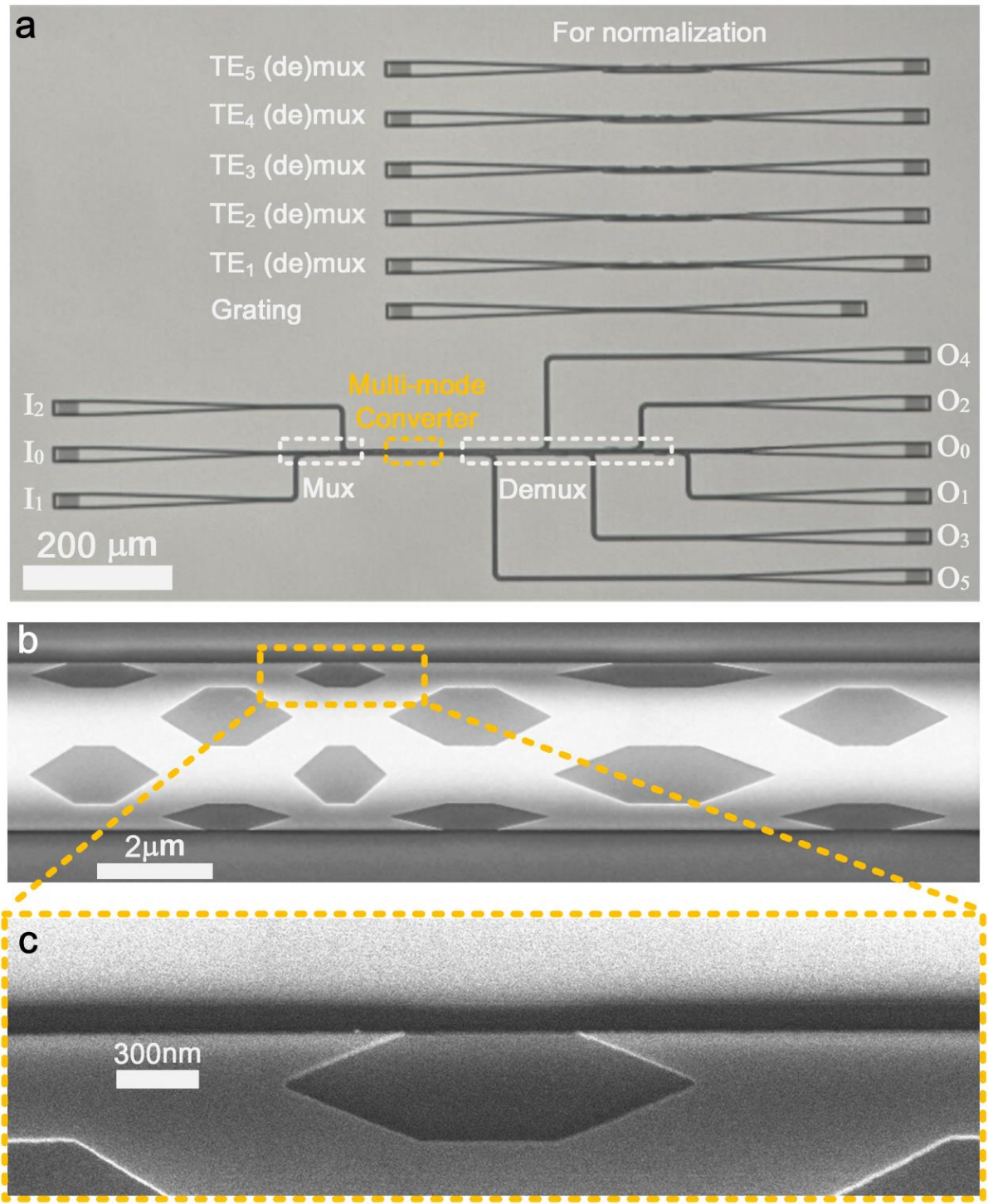

**Figure 4 | Photos of the fabricated multi-mode converter.** (**a**) Optical Microscope and (**b**) SEM image of the fabricated devices. (**c**) Magnified SEM image of the hexagonal trenches.

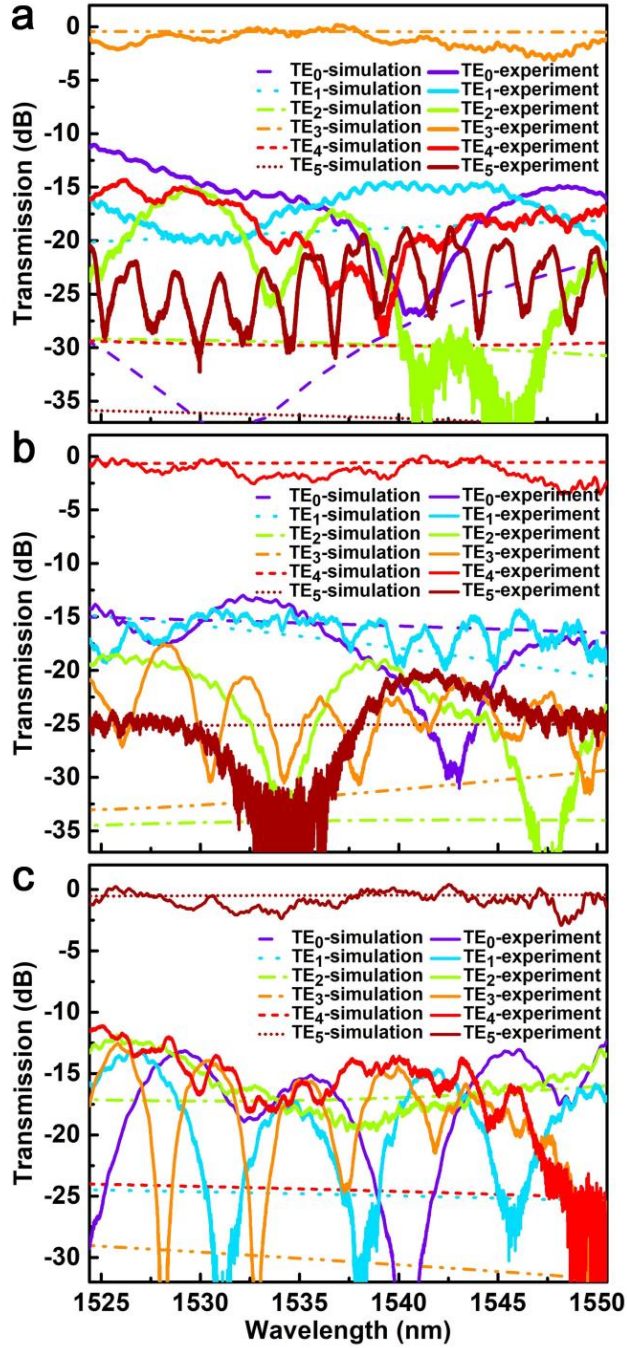

**Figure 5 | Measurement results of the fabricated multi-mode converter.** Measured transmission spectra of the fabricated $TE_i$-to-$TE_{i+3}$ ($i$ = 0, 1, 2) multi-mode converter for the (**a**) $TE_0$, (**b**) $TE_1$ and (**c**) $TE_2$ mode input, respectively. Simulated transmission spectra are also provided for comparison, as shown by the dashed curves.

**Tables**

**Table 1 | Detailed design parameters for the $TE_i$-to-$TE_{i+3}$ ($i$ = 0, 1, 2) multi-mode converter with hexagonal trenches**

| Width<br>($W_1$ ~ $W_4$) [μm] | Length<br>($T_1$ ~ $T_6$) [μm] | Length<br>($L_1$ ~ $L_6$) [μm] |
|---|---|---|
| $W_1$ = 0.46, $W_2$ = 1.04 | $T_1$ = 0.81, $T_2$ = 0.83 | $L_1$ = 0.69, $L_2$ = 0.71 |
| $W_3$ = 1.04, $W_4$ = 0.46 | $T_3$ = 0.58, $T_4$ = 1.02 | $L_3$ = 0.50, $L_4$ = 0.87 |
|  | $T_5$ = 1.39, $T_6$ = 1.06 | $L_5$ = 1.12, $L_6$ = 0.91 |

The height and width of the silicon waveguide are 0.22 μm and 3 μm, respectively; the length of the metastructure is 16.2 μm; the etching depth is 0.05μm.

# Supplementary Information

# Multi-mode conversion via two-dimensional refractive-index perturbation on a silicon waveguide

Chunhui Yao, Zhen Wang, Hongwei Wang, Yu He, Yong Zhang & Yikai Su*

State Key Laboratory of Advanced Optical Communication Systems and Networks, Department of Electronic Engineering, Shanghai Jiao Tong University, Shanghai 200240, China

**Supplementary Figures**

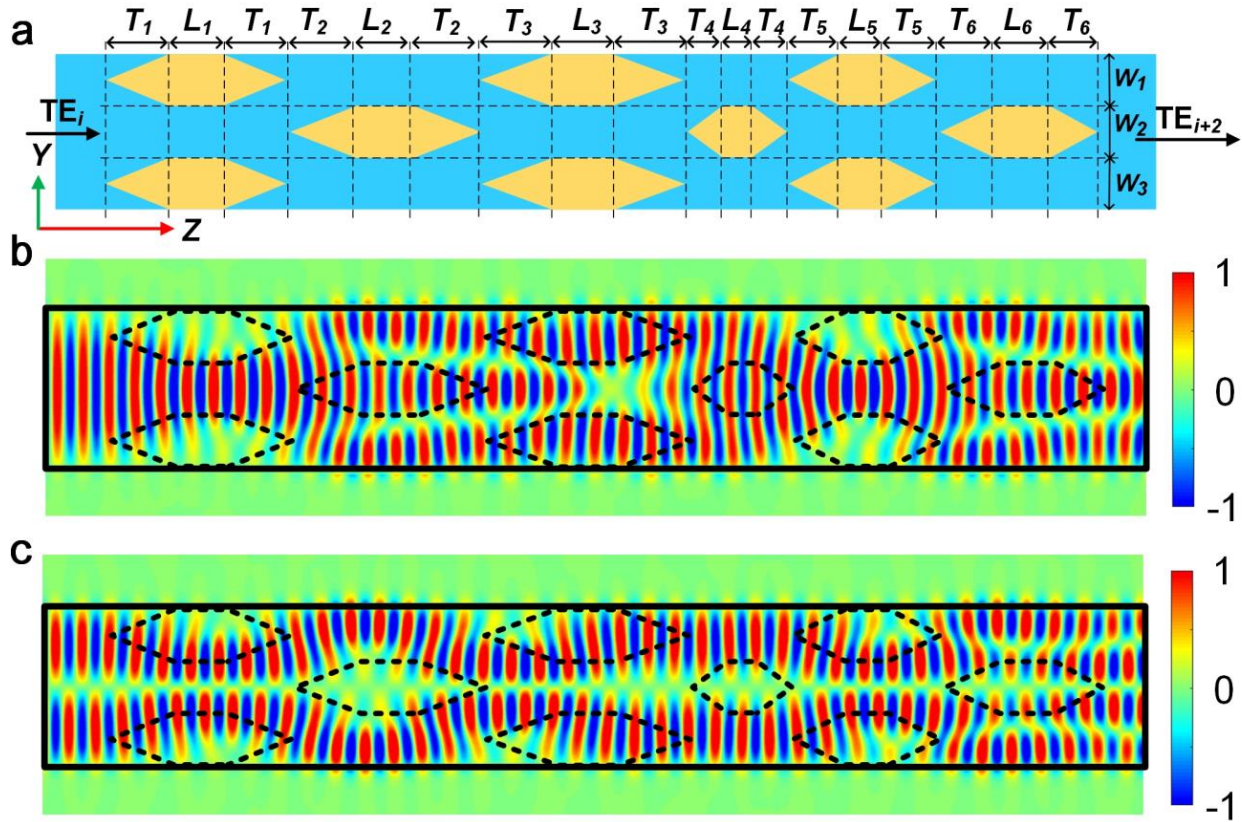

**Supplementary Figure 1 | Simulation results of the TE$_i$-to-TE$_{i+2}$ ($i$ = 0, 1) multi-mode converter.** (**a**) Top view of the proposed TE$_i$-to-TE$_{i+2}$ ($i$ = 0, 1) multi-mode converter with hexagonal trenches. Simulated electric field ($E_y$) distribution for the (**b**) TE$_0$, (**c**) TE$_1$ mode input, respectively.

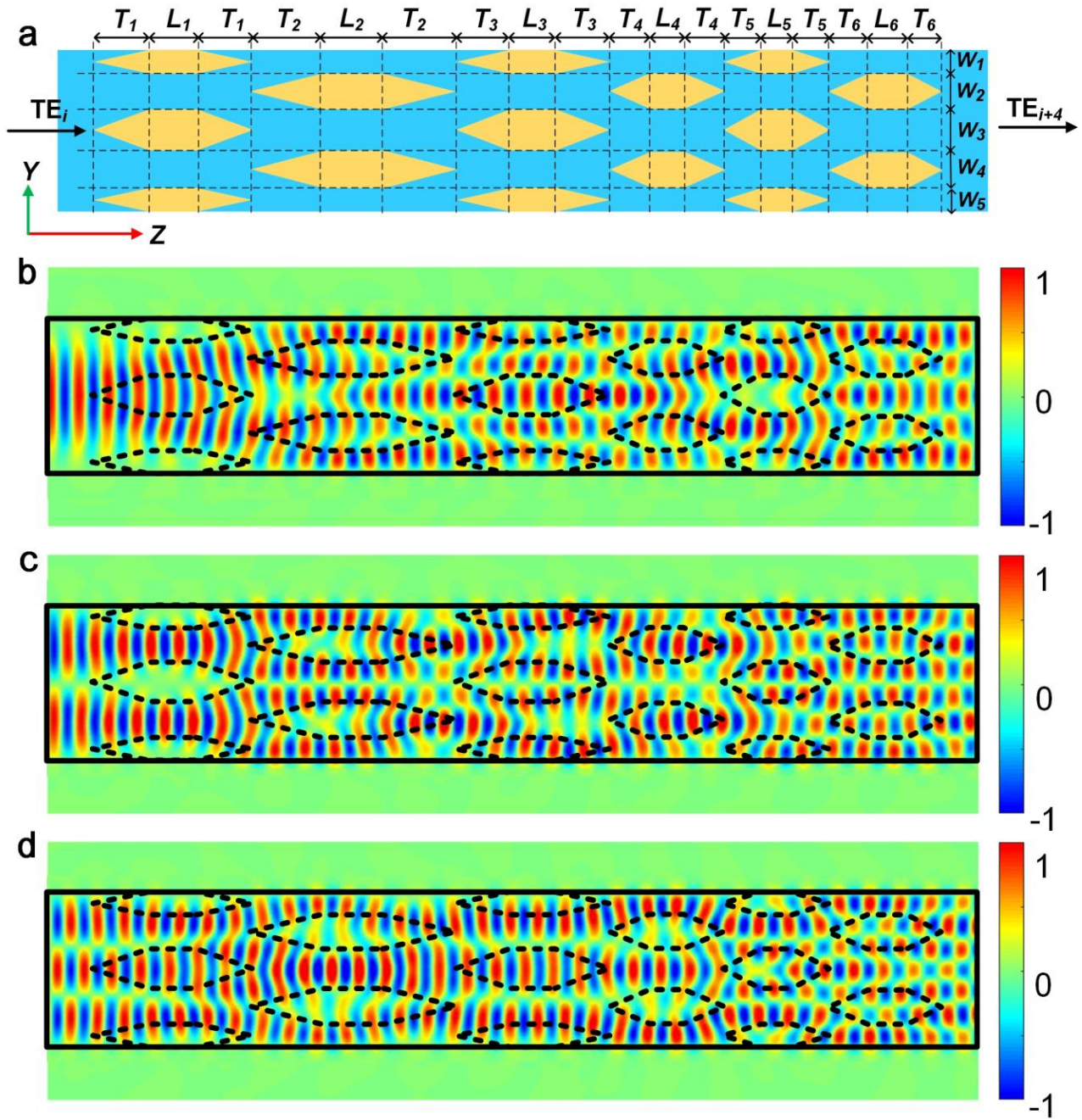

**Supplementary Figure 2 | Simulation results of the TE$_i$-to-TE$_{i+4}$ ($i$ = 0, 1, 2) multi-mode converter.** (**a**) Top view of the proposed TE$_i$-to-TE$_{i+4}$ ($i$ = 0, 1, 2) multi-mode converter with hexagonal trenches. Simulated electric field ($E_y$) distribution for the (**b**) TE$_0$, (**c**) TE$_1$, (**d**) TE$_2$ mode input, respectively.

## Supplementary Tables

**Supplementary Table 1 | Detailed design parameters and simulation results of the TE$_i$-to-TE$_{i+3}$ ($i$ = 0, 1, 2) multi-mode converter with rectangular trenches**

| Width ($W_1$ ~ $W_4$) [μm] | Length ($L_1$ ~ $L_6$) [μm] | Insertion losses [dB] | Crosstalk [dB] |
|---|---|---|---|
| $W_1$=0.46, $W_2$=1.04 | $L_1$=2.40, $L_2$=2.20 | TE$_0$-to-TE$_3$: 0.99 | −15.5 |
| $W_3$=1.04, $W_4$=0.46 | $L_3$=1.75, $L_4$=2.80 | TE$_1$-to-TE$_4$: 0.97 | −13.4 |
| | $L_5$=3.90, $L_6$=2.85 | TE$_2$-to-TE$_5$: 1.02 | −17.2 |

The height and width of the silicon waveguide are 0.22 μm and 3 μm, respectively; the length of the metastructure is 15.9 μm; the etching depth is 0.05 μm.
.

**Supplementary Table 2 | Detailed design parameters and simulation results of the TE$_i$-to-TE$_{i+2}$ ($i$ = 0, 1) multi-mode converter with hexagonal trenches**

| Width ($W_1$ ~ $W_3$) [μm] | Length ($T_1$ ~ $T_6$) [μm] | Length ($L_1$ ~ $L_6$) [μm] | Insertion losses [dB] | Crosstalk [dB] |
|---|---|---|---|---|
| $W_1$=0.68, | $T_1$=1.31, $T_2$=1.42 | $L_1$=1.28, $L_2$=1.39 | TE$_0$-to-TE$_2$: 0.48 | −15.2 |
| $W_2$=1.04 | $T_3$=1.42, $T_4$=0.67 | $L_3$=1.39, $L_4$=0.66 | TE$_1$-to-TE$_3$: 0.44 | −12.8 |
| $W_3$=0.68 | $T_5$=1.07, $T_6$=1.14 | $L_5$=1.06, $L_6$=1.12 | | |

The height and width of the silicon waveguide are 0.22 μm and 2.4 μm, respectively; the length of the metastructure is 20.9 μm; the etching depth is 0.05 μm.
.

**Supplementary Table 3 | Detailed design parameters and simulation results of the $TE_i$-to-$TE_{i+4}$ ($i = 0, 1, 2$) multi-mode converter with hexagonal trenches**

| Width ($W_1 \sim W_5$) [μm] | Length ($T_1 \sim T_6$) [μm] | Length ($L_1 \sim L_6$) [μm] | Insertion losses [dB] | Crosstalk [dB] |
|---|---|---|---|---|
| $W_1$=0.53, $W_2$=0.80 | $T_1$=0.99, $T_2$=1.29 | $L_1$=1.01, $L_2$=1.32 | $TE_0$-to-$TE_4$: 0.54 | −14.4 |
| $W_3$=0.95, $W_4$=0.80 | $T_3$=0.92, $T_4$=0.73 | $L_3$=0.96, $L_4$=0.74 | $TE_1$-to-$TE_5$: 0.71 | −16.0 |
| $W_5$=0.53 | $T_5$=0.63, $T_6$=0.69 | $L_5$=0.64, $L_6$=0.72 | $TE_2$-to-$TE_6$: 0.67 | −12.4 |

The height and width of the silicon waveguide are 0.22 μm and 3.6 μm, respectively; the length of the metastructure is 15.8 μm; the etching depth is 0.055 μm.

**Supplementary Table 4 | Detailed design parameters for the mode (de)multiplexers**

| Mode (de) Multiplexers | Width of bus waveguide[μm] | Width of access waveguide[μm] | Gap [μm] | Coupling length[μm] |
|---|---|---|---|---|
| $TE_1$ | 0.81 | 0.4 | 0.1 | 11.5 |
| $TE_2$ | 1.22 | 0.4 | 0.1 | 14 |
| $TE_3$ | 1.63 | 0.4 | 0.1 | 16.5 |
| $TE_4$ | 2.05 | 0.4 | 0.1 | 18 |
| $TE_5$ | 2.46 | 0.4 | 0.1 | 21 |